# What does photon energy tell us about cellphone safety?

William J. Bruno, Ph.D.

Theoretical Biology & Biophysics Los Alamos National Laboratory

Current address: New Mexico Consortium, Los Alamos, NM

April 24, 2011; latest revision April 25, 2017

**Abstract**

It has been argued that cellphones are safe because a single microwave photon does not have enough energy to break a chemical bond. We show that cellphone technology operates in the classical wave limit, not the single photon limit. Based on energy densities relative to thermal energy, we estimate thresholds at which effects could be possible. These seem to correspond somewhat with many experimental observations.

It has been argued repeatedly[Park 2001, 2002, 2006, 2009, 2010, 2011, Shermer 2010] that cellphones must be safe because a single microwave photon does not have enough energy to break a chemical bond. This argument would perhaps be convincing if the photon flux were less than 1 photon per square wavelength per photon period (equivalent to a photon density of < 1 per cubic wavelength). However, this condition, which holds for some common sources of ionizing radiation, does not hold for cellphone exposures (Table 1). This means that while ionizing radiation is typically in the pure quantum limit of low photon density, cellphones and cell towers operate in the classical wave limit of high photon densities. In this situation the energy of each photon is often irrelevant.

Table 1:

| Source | Approximate photon density per cubic wavelength |
|---|---|
| Medical X-ray | ~ 1e-24 |
| Sunlight UV | ~ 1e-7 |
| Cell tower (~10 meters away) | ~ 1e+15 |
| Cellphone | ~ 1e+20 |

Notes: Microwaves assumed ~1GHz; cellphone ~300V/m; Cell tower ~1V/m. Sunlight ~10W/m^2, 300nm. X-ray: 30 cm from 1mA source, 1% efficient.

That coherent photon energies can combine to do work (including work other than just heating) is most clearly illustrated by optical tweezers, which can be used to move bacterial cells but cause physiological damage in the process [Rasmussen et al. 2008]. The requirements for biological tweezers to operate are a gradient in the index of refraction and sufficient flux of photons (proportional to the work to be done). Table 1 indicates a large flux of photons, the energy content of which we analyze below. Gradients in refractive index are present at every membrane/cytosol (or nucleosol) interface as well as at edges of myelin sheath or any subcellular structure, ultrastructure or vesicle. In fact, non-thermal microwave damage to ultrastructure has been reported [Webber et al., 1980], and there are many reports of cellphone signals damaging the blood-brain barrier (e.g., Salford et al. 2003). Because of the importance of this barrier (e.g., for protecting glutamergic neurons from glutamate; it is primarily these neurons that are progressively lost in Alzheimer's disease) such damage could be expected to lead to multiple harmful effects.

Another example of how an optical tweezer-like effect might come about is microwave hearing. Sharp et al. [1974] proposed photon pressure as the mechanism for this well established effect, and also for the observation that objects like crumpled foil or paper emit sound when exposed to strong, but non-thermal, pulsed microwaves.

Another established effect in which photon energies combine to apply a force is "pearl chain formation", in which colloidal or other particles are forced into alignment by an RF field. This effect is clearly analogous to the rouleaux formation reported by Havas [2010]. There is a literature claiming that pearl chain formation only happens when the fields are strong enough to cause significant thermal heating, but obviously this would depend on the relative values of the real and imaginary permittivities, which vary with tissue and frequency.

Surely there must be some safe level of microwave flux below which we can rule out effects on the basis of physical arguments. Levels well below the natural microwave background (mainly from the sun) would not be noticed (at least during the day). Unfortunately, this level is very low by cellphone-technology standards, some 8 to 9 orders of magnitude lower than common cell tower exposures. More modestly one might expect that in the absence of any sharp resonances or large focusing effects, a level on the order of the average thermal energy, $k_B T$, per cubic wavelength should be safe. This would correspond to about 30pW/m^2 (at ~1 GHz), again very low. This equates to exposure from a cell tower at a distance of a few miles. That is on the same scale as the threshold at which Bise (1978) reported changes to human EEG. (Incidentally, the Bise experiments were dismissed in a review by industry-oriented scientists [D'Andrea et al. 2003], on the basis that the effects are seen below urban "background" levels. However, the background levels referred to are actually mainly from FM radio broadcast at ~100 MHz, which is much less efficient at entering the brain [Frey 1962].) We now know that the EEG affects neural firing [Anastassiou et al., 2011]. Headaches [Hutter 2006] and a number of other effects [Santini 2003, Eger, 2010] including sleep loss and depression have been reported in people living at various distances near cell towers. Cell tower level effects have also been observed on bees [Sharma et al., 2010] and frogs [Balmori 2010].

To be still less cautious, we could hope that if the energy present over a cell volume is less than $k_B T$, then there should be no damage at the cellular level. In fact, biological structures must have a stability of at least several $k_B T$, suggesting short term exposures will have an extra margin of safety. Long term exposures of just over 1 $k_B T$ would be expected to marginally accelerate any existing aging processes (the emerging understanding of neurodegenerative disease is that repair processes cannot keep up with the rate of molecular damage to the neuron [Martinez-Vicente & Cuervo 2007].

Limiting the level of exposure on the basis of a single cell is only likely to go wrong if there are multicellular structures that concentrate RF energy from a larger volume into one cell. This could happen due to resonances, or focusing, or conductive 'circuits' (the presence of apparent semiconductors such as neuromelanin and biogenic calcite in the brain, and of piezo-electric collagen, should inspire more research into whether such circuits exist). Nevertheless, we compute a safety ballpark level using this approach of 1000 V/m for small (10 micron diameter) cells. For a very large neuron (100 micron diameter) a safe exposure would be only 30 V/m, which is less than the hundreds of V/m a cellphone typically emits. Note that the human body contains a wide range of neuron sizes (up to ~1 meter long), and that both in normal aging and more so in Alzheimer's disease, there is a progressive decrease in the number of large neurons in the brain [Terry et al., 1987].

Many effects have been reported from cellphone level exposures. These include sleep disruption [Lowden et al., 2011], changes in brain metabolism that persist at least 5 minutes after use [Volkow et al., 2011], increased risk of tinnitus [Hutter et al., 2010], and increased risk of brain tumors [e.g., Myong et al., 2009] and salivary gland tumors, in addition to the previously mentioned animal studies finding damage to the blood-brain barrier. For phones worn on the hip, studies finding sperm damage [De Iuliis 2009] and hip bone density asymmetry [Saravi 2011] have also been published. Based on the physics and biology described here and elsewhere [Hyland 2000], it is not implausible that such effects could be real. In fact, it could be argued that the supposed absence of any harmful effects would be a more surprising, though more welcome, outcome. Indeed although the best quality epidemiological studies (reviewed by Myong et al. 2009) see increased tumors, many other studies have failed to observe effects. Thorough analyses of the negative experiments shows that in many cases they are actually compatible with the positive findings [Morgan 2009, Slesin 2010].

Mobile communications have been proven to be of tremendous value and popularity. The current approach to dosimetry, evidently modeled on that used for ionizing radiation, appears to be broken, and in fact has been criticized essentially since its inception [e.g., Frey 1994; Gandhi, 1987]. Arguments in support of safety based on basic physics appear not to hold up.

The current technology is far from optimal in terms of biological compatibility, considering that microwaves in the 1-10 GHz frequency range most efficiently do work inside the brain [Frey 1962], and current digital pulse modulation schemes makes use of frequencies that, if demodulated [Bruno 2011], are also used by neurons. Frequencies above 10 GHz deposit most of their energy in the skin, while lower frequencies (traditional TV and radio) are thought to be reflected without much transfer of energy [Frey 1962]. Visible light, in the form of through-space optical wireless, may offer hope of high bandwidth wireless (though limited to line-of-sight) and the possibility of long-term safety, although careful consideration of visible light's role in regulatory pathways (including vitamin D and melatonin) would still be required.

**References**

Anastassiou, C.A., Rodrigo Perin, Henry Markram & Christof Koch, Ephaptic coupling of cortical neurons, Nature Neuroscience 14: 217–223 (2011)

Balmori A, Mobile phone mast effects on common frog (Rana temporaria) tadpoles: the city turned into a laboratory. Electromagn Biol Med 2010; 29 (1-2): 31 - 35

Bruno, William J. A predicted Mechanism for Biological Effects of Radio-Frequency Electro-Magnetic Fields: Piezoelectric Rectification, Biophysical Journal Volume 100, Issue 3, Supplement 1, 2 February 2011, Page 623a

D'Andrea JA, Chou CK, Johnston SA, Adair ER. Microwave effects on the nervous system. Bioelectromagnetics. 2003;Suppl 6:S107-47.

De Iuliis GN, Newey RJ, King BV, Aitken RJ, 2009 Mobile Phone Radiation Induces Reactive Oxygen Species Production and DNA Damage in Human Spermatozoa In Vitro. PLoS ONE 4(7): e6446

Eger, Horst and Manfred Jahn Specific Health Symptoms and Cell Phone Radiation in Selbitz (Bavaria, Germany)— Evidence of a Dose-Response Relationship

umwelt-medizin-gesellschaft 23, 2/2010
Frey, Allan H., Human auditory system response to modulated electromagnetic energy. J Appl

Physiol. 1962 Jul;17:689-92.

Frey, Allan H. (ed) (1994). On the Nature of Electromagnetic Field Interactions with Biological Systems. Austin, TX: R.G. Landes Company, (pages 5-6).

Gandhi, OP, The ANSI radio frequency safety standard: Its rationale and some problems. IEEE Engineering in Medicine and Biology [IEEE ENG. MED. BIOL.]. Vol. 6, no. 1, pp. 22-25. 1987.

Havas, Magda, "Live blood and electrosmog". http://www.magdahavas.com/2010/03/22/live-blood- cells-and-electrosmog/ (2010).

Hutter, H-P, H Moshammer, P Wallner and M Kundi. 2006. Subjective symptoms, sleeping problems, and cognitive performance in subjects living near mobile phone base stations. Occup.

Environ. Med 63;307-313 .

Hutter et al. Tinnitus and mobile phone use, Occup Environ Med 2010;67:804-808

Hyland, G.J. The physics and biology of mobile telephony. The Lancet 356, 1833-1836 (2000).

Morgan, Lloyd, Cell Phones and Brain Tumors: 15 Reasons for Concern," The Radiation Research Trust, August 25, (2009).

Park, Robert L. Cellular Telephones and Cancer: How Should Science Respond? NCI J Natl Cancer Inst 93 (3): 166-167 (2001)

Park, Robert, "What's New", Oct 4 (2002) Dec 8, (2006) Sep 25, Oct 30, Dec 25, (2009) Jan 8, Mar 12,Mar19,Apr23,May21,Jun4,Jul2, Nov5,Dec10,Dec17 (2010),Jan1,Feb4,Feb25, (2011).

Salford LG, Brun AE, Eberhardt JL, Malmgren L, Persson BR. Nerve cell damage in mammalian brain after exposure to microwaves from GSM mobile phones. Environ Health Perspect. 2003 Jun; 111(7):881-3;

Santini, R; Santini, P; Le Ruz, P; Danze, JM; Seigne, M (2003). Survey study of people living in the vicinity of cellular phone base stations, Electromagnetic Biology and Medicine, 22 (1): 41-49

Saravi FD, Asymmetries in hip mineralization in mobile cellular phone users. J Craniofac Surg. 2011 Mar;22(2):706-10.

Sharma V. P and Kumar Neelima , Changes in honeybee behaviour and biology under the influence of cellphone radiation, Current Science (India), Vol. 98, No. 10, 25 2010

Sharp, J.C.; Grove, H.M.; Gandhi, O.P.; Generation of Acoustic Signals by Pulsed Microwave Energy Microwave Theory and Techniques, IEEE Transactions on, 22 (5): 583 - 584 (1974).

Shermer M., Skeptic: Can You Hear Me Now? The Truth about Cellphones and Cancer Scientific American, Oct, p. 98 (2010)

Myung et al., 2009, Mobile Phone Use and Risk of Tumors: A Meta-Analysis. Journal of Clinical Oncology, 27(33):5565-72.

Lowden A, et al. Sleep after mobile phone exposure in subjects with mobile phone-related symptoms. Bioelectromagnetics. 2011 Jan;32(1):4-14.

Martinez-Vicente, M, Cuervo M, Autophagy and neurodegeneration: when the cleaning crew goes on strike, The Lancet Neurology, Volume 6, Issue 4, Pages 352 - 361, April 2007

Rasmussen et al (2008) Appl Environ Microbiol. 2008 April; 74(8): 2441
Slesin, Louis, Interphone Points to Long-Term Brain Tumor Risks, MicrowaveNews, May 17, 2010 RD Terry, R DeTeresa & LA Hansen. Ann Neurol. 1987, 21:532

Volkow, Nora D. , Dardo Tomasi, Gene-Jack Wang, Paul Vaska, Joanna S. Fowler, Frank Telang, Dave Alexoff, Jean Logan, Christopher Wong, Effects of Cell Phone Radiofrequency Signal Exposure on Brain Glucose Metabolism JAMA. 2011;305(8):808-813.

Webber, Mukta M., Frank S. Barnes, Linda A. Seltzer, Thomas R. Bouldin, and Kedar N. Prasad. Short microwave pulses cause ultrastructural membrane damage in neuroblastoma cells. Journal of Ultrastructure Research Volume 71, Issue 3, June 1980, Pages 321-330

## Response to Bernard Leikind

The most salient argument in the lengthy the criticism (over twice as long as our paper!) by independent physicist Bernard Leikind [1] is that when many degrees of freedom are present, it seems unlikely that any one of them will absorb a significant fraction of the available energy. This might be true in other contexts, but it is well known that microwave energy at cell phone frequencies penetrates deeply into tissue (decay length > 2 cm;  hence a microwave oven heats meat "from the inside out"). This means almost all degrees of freedom remove almost no energy, and their existence becomes unimportant. Therefore, there remains enough energy for bioeffects from cell phones to be physically possible, just as it is physically possible for a cellphone to receive its signal. Note that the phone uses more energy from its battery when a call comes in; hence the incoming wave ultimately caused the breaking and rearrangement of chemical bonds in the battery. Leikind seems to agree with our main point that these signals are carried by many-photon fields that should be analyzed in the classical limit, although he had previously used the single photon argument [2].  Our conclusion remains: that earlier, simplistic quantum arguments should be rejected, especially in the face of the growing experimental literature showing that such bioeffects do exist.

**Additional References**

[1] Leikind, Bernard. "An Assessment of" What does photon energy tell us about cellphone safety" by Dr. William Bruno." *arXiv preprint arXiv:1107.0086* (2011).

[2] Leikind, Bernard, "Do Cell Phones Cause Cancer?" Skeptic 15, No. 4, pp. 30-34, (2010).